\documentclass[a4paper]{jpconf}

\usepackage{mathtools}
\usepackage{amssymb}
\usepackage[hidelinks]{hyperref}
\usepackage[capitalise]{cleveref}
\usepackage[square,sort&compress,numbers]{natbib}
\usepackage{graphicx}
\usepackage{xspace}
\usepackage[nolist,printonlyreused]{acronym}
\usepackage{siunitx}

\graphicspath{{./img/}}

\usepackage[capitalise]{cleveref}
\makeatletter
\newcommand{\crefext}[2]{\csname cref@#1@format\endcsname{#2}{}{}}
\newcommand{\Crefext}[2]{\csname Cref@#1@format\endcsname{#2}{}{}}
\newcommand{\crefextp}[2]{\csname cref@#1@name@plural\endcsname{~#2}{}{}}
\newcommand{\Crefextp}[2]{\csname Cref@#1@name@plural\endcsname{~#2}{}{}}
\makeatother

\newcommand{\code}[1]{\texttt{#1}\xspace}

\newcommand{\cpp}{\code{C++}}
\newcommand{\eigen}{\code{Eigen}}
\newcommand{\finiteflow}{\code{FiniteFlow}}

\newcommand{\gcc}{\code{GCC}}

\newcommand{\njet}{\code{NJet}}
\newcommand{\nnlojet}{\code{NNLOjet}}
\newcommand{\openloops}{\code{OpenLoops2}}
\newcommand{\pentagonfunctions}{\code{PentagonFunctions++}}

\newcommand{\qd}{\code{QD}}

\newcommand{\incite}[1]{Ref.~\cite{#1}}

\begin{document}

\begin{acronym}
    \acro{CPU}{central processing unit}
    \acro{LO}{leading order}
    \acro{N3LO}[N\textsuperscript{3}\acs*{LO}]{(next-to-)\textsuperscript{3}leading order}
    \acro{NLO}{next-to-leading order}
    \acro{NNLO}[NN\acs*{LO}]{next-to-next-to-leading order}
    \acro{QCD}{quantum chromodynamics}
    \acro{f128}[\texttt{f128}]{128-bit floating-point number}
    \acro{f256}[\texttt{f256}]{256-bit floating-point number}
    \acro{f32}[\texttt{f32}]{32-bit floating-point number}
    \acro{f64}[\texttt{f64}]{64-bit floating-point number}
\end{acronym}

\title{Two-loop amplitudes in massless QCD for diphoton-plus-jet production via gluon fusion at hadron colliders}
\author{Ryan Moodie}
\address{
    Dipartimento di Fisica and Arnold-Regge Center, Università di Torino, and INFN, Sezione di Torino, Via P. Giuria 1, I-10125 Torino, Italy
}
\ead{ryaniain.moodie@unito.it}

\begin{abstract}
    The ability to calculate analytic full-colour two-to-three two-loop helicity amplitudes in massless \acs{QCD} has been a recent triumph of the field, driving phenomenology towards one percent precision.
    In this contribution, we focus on the virtual correction of diphoton-plus-jet production via gluon fusion, $gg\to g\gamma\gamma$.
    We outline state-of-the-art strategies to optimise the reconstruction over finite fields of the rational coefficients of the amplitude.
    We also present an updated performance analysis of our publicly-available \cpp implementation of the amplitude, demonstrating its typical speed and stability, and study its stability in infrared regions of phase space.
    These results are relevant for improving predictions in Higgs physics at hadron colliders.
\end{abstract}

\section{Introduction}

Precise theoretical predictions are in high demand for the current Large Hadron Collider experiments, which aim to better understand the properties of the Standard Model while indirectly probing for new physics through tiny deviations.
With experimental bottlenecks like the determination of interaction luminosity at around one percent~\cite{!Atlas:2019pzw,Cms:2021xjt} and similarly for the resolution of jet energies~\cite{Cms:2016lmd,Atlas:2017bje}, the current target for theory is to also achieve one percent precision~\cite{Salam:2018rwo}.
In the near future, the High Luminosity upgrade will also overcome statistical limitations~\cite{!Dainese:2019rgk}.
Fixed-order matrix elements are one of several ingredients in theoretical predictions requiring improvement to achieve this goal.

Due to the relatively large size of the strong coupling constant $\alpha_s$, \ac{NNLO} corrections in \ac{QCD} are desirable for a wide variety of processes.
In particular, the challenge of two-to-three scattering processes has been met with the development of new methods capable of overcoming their algebraic and analytic complexity.
We are now seeing the first calculations of such amplitudes and distributions in massless and single-external-mass configurations~\cite{!Czakon:2021mjy,Chawdhry:2021hkp,Badger:2021imn,Chawdhry:2019bji,Sotnikov:2022wlx}.

Diphoton production is an important experimental signature at hadron colliders~\cite{Amoroso:2020lgh} and can be used to study the Higgs boson through its decay to photons.
Diphoton-plus-jet signatures, $pp\to j\gamma\gamma$, form the largest background to Higgs production at high transverse momenta.
Leading-colour \ac{NNLO} distributions~\cite{Chawdhry:2021hkp} and full-colour two-loop amplitudes~\cite{Agarwal:2021vdh} for this process have been calculated.
The distributions display good perturbative convergence except in regions where the gluon-fusion subprocess, first appearing at \ac{NNLO}, is relatively large.
Our calculation of the full-colour two-loop $gg\to g\gamma\gamma$ amplitudes~\cite{Badger:2021imn} contribute at \ac{NLO} in the gluon-fusion subprocess and have been used for full-colour distributions~\cite{Badger:2021ohm} at this order.
Their inclusion in the full process calculation offers to tame the scale uncertainties in these regions of phase space.

\section{Computation of analytic expression}

Our aim is to obtain analytic expressions for two-loop helicity amplitudes.
We organise the colour-ordered amplitudes, or rather their finite remainders $F$ after infrared subtraction, as
\begin{align}
    \label{eq:decom}
    F(\boldsymbol{x}) = \sum_i r_i(\boldsymbol{x}) f_i(\boldsymbol{x}),
\end{align}
where $\boldsymbol{x}$ denotes the set of independent variables in some kinematic parametrisation, $r_i$ are rational coefficients, and $f_i$ are special functions resulting from the integrals.
We use momentum twistor variables for $\boldsymbol{x}$~\cite{Hodges:2009hk} and monomials of the pentagon functions for $f_i$~\cite{Chicherin:2020oor,Chicherin:2021dyp}.

In our workflow~\cite{!Krys:2022gby,Badger:2021imn}, we perform the bulk of the computation numerically to bypass intermediate complexity of analytic expressions.
We evaluate using finite field arithmetic~\cite{Peraro:2016wsq} to avoid the precision-loss problems of floating-point numbers.
We start from Feynman diagrams and process to a form where the integral coefficients can be loaded into a \finiteflow~\cite{Peraro:2019svx} dataflow graph.
Further manipulation, including coefficient mappings, integration-by-parts reduction, projection to the pentagon function basis, Laurent expansion in the dimensional regulator, and finally multivariate reconstruction of the $r_i$ from numerical evaluations over finite fields, proceed within \finiteflow.
We obtain compact analytic forms of $r_i$ as output, which can be implemented in libraries for efficient evaluation as in \cref{sec:impl}.

The full details of the $gg\to g\gamma\gamma$ reconstruction are presented in \incite{Badger:2021imn}.
Here, we highlight some optimisation strategies used to speed up the reconstruction by reducing the required number of sample points.

\subsection{Linear relations in the coefficients}

The representation \cref{eq:decom} is suboptimal as there are linear relations between the $r_i$.
Expressing in terms of a set of linearly independent $r_i$ simplifies the reconstruction.
We can determine the linear relations by solving the linear fit,
\begin{align}
    \sum_i a_i \, r_i(\boldsymbol{x})  = 0 \, .
\end{align}
When choosing the linearly independent subset, preferring simpler $r_i$ further optimises the reconstruction.
We use polynomial degrees to estimate the complexity of expressions.

\subsection{Matching factors in the denominator}
\label{sec:matching-factors}

The pole structure of the pentagon functions is determined by the letters of the pentagon alphabet, $\{\ell_k\}$~\cite{Chicherin:2017dob}.
Expecting the poles of the rational coefficients to be linked to those of the special functions they multiply, we make for each coefficient the ansatz,
\begin{align}
    \label{eq:ansatz}
    r(\boldsymbol{x}) &= \frac{n(\boldsymbol{x})}{\prod_{k} {\ell_k}^{e_k}(\boldsymbol{x})}\,,
\end{align}
where $e_k$ are integers and $n(x)$ is a polynomial in the variables $\boldsymbol{x}$.
We determine the $e_k$ by reconstructing $r$ on a univariate slice~\cite{Abreu:2018zmy}, defined by parametrising the variables $\boldsymbol{x}$ by a single parameter $t$ as $\boldsymbol{x}(t) = \boldsymbol{c}_0 + \boldsymbol{c}_1 t$, with vectors $\boldsymbol{c}_i$ constant and randomly assigned, and matching the reconstructed $r(t)\coloneqq r(\boldsymbol{x}(t))$ with \cref{eq:ansatz} evaluated on the same slice.
This entirely fixes the denominator, as well as some factors of the numerator appearing with negative $e_k$, simplifying the reconstruction.

\subsection{Univariate partial fraction decomposition}

Partial fractioning of rational functions can yield more compact expressions.
To simplify the reconstruction, we reconstruct the coefficients in a form which is decomposed in univariate partial fractions.

Let us consider the partial fractioning of a rational function with respect to the variable $y$.
As discussed in \cref{sec:matching-factors}, we can infer the $y$-dependent part of the denominator.
For example, we could obtain, with its parametrised decomposition,
\begin{align}
    \label{eq:pfd}
    r(x,y) = \frac{n(x,y)}{y^2(x^2+y^2)} = \frac{q_1(x)}{y} + \frac{q_2(x)+q_3(x)y}{y^2}+\frac{q_4(x)+q_5(x)y}{x^2+y^2} + \sum_{i=0}^{d-4}q_{6+i}(x) \, y^{i}.
\end{align}
The undetermined part of $r(x,y)$ is $n(x,y)$, which has degree $d$ in $y$.
To determine $d$, we reconstruct $n(x,y)$ on a univariate slice varying only $y$, $\{x(t) = c,\, y(t) = t\}$, with $c$ constant, such that $d$ is given by the degree in $t$ of $n(t)$.
The $4$ in the sum of \cref{eq:pfd} is the maximal degree in $y$ of the denominator of $r(x,y)$.
We then reconstruct the $q_i(x)$ using a linear fit in \finiteflow over numerical evaluations of $r(x,y)$.
Since the fit involves several evaluations, each sample of $\{q_i(x)\}$ in the reconstruction is more expensive than that of $r(x,y)$, but depends on one fewer variable and has substantially lower degrees, so fewer samples are required.
Therefore, the partial fractioned reconstruction can outperform the direct one, particularly for complex functions such as those appearing at two loops.
The choice of $y$ is crucial; testing at lower loop orders can quickly inform this decision.

\section{Library implementation}
\label{sec:impl}

The $\ell$-loop finite remainders $\mathcal{F}^{(\ell)}$ are implemented in the \njet \cpp library~\cite{!njet}, which is linked to the \pentagonfunctions library~\cite{Chicherin:2021dyp} for the evaluation of the special functions.
The decomposition of the finite remainders is detailed in \incite{Badger:2021imn}.
The code returns the value and error, estimated using the dimension scaling test, for the colour- and helicity-summed (denoted by $\otimes$) hard functions, $\mathcal{H}^{(1)}$ and $\mathcal{H}^{(2)}$, 
\begin{align}
    \begin{aligned}
        \mathcal{H} &= \frac{\alpha^2 \alpha_s^3}{(4\pi)^5} \left(
        \mathcal{H}^{(1)} + \frac{\alpha_s}{4\pi} \, \mathcal{H}^{(2)}
        \right) + \mathcal{O}(\alpha_s^5),\\
        \mathcal{H}^{(1)} &= \mathcal{F}^{(1)} \otimes \mathcal{F}^{(1)},\\
        \mathcal{H}^{(2)} &= \mathcal{F}^{(2)} \otimes \mathcal{F}^{(1)}.
    \end{aligned}
\end{align}

We provide an evaluation strategy that reevaluates the result with higher precision floating-point numbers if a user-provided target accuracy is not achieved.
The code is templated so that we can use native \acp{f64} and \acp{f128} from \qd~\cite{!qdweb} to evaluate the rational coefficients and special functions with independent precisions.
Labelling the coefficients first and the special functions second, we define three tiers: \ac{f64}/\ac{f64}, \ac{f128}/\ac{f64}, and \ac{f128}/\ac{f128}.

\section{Performance}
\label{sec:performance}
% data generation: https://gitlab.com/jetdynamics/njet-tools/-/blob/master/3g2a%402l/test/collab/calc2.cpp
% data cached: https://gitlab.com/jetdynamics/njet-tools/-/tree/master/3g2a%402l/test/collab/3g2a_phys_res
% plot: https://gitlab.com/jetdynamics/njet-tools/-/blob/master/3g2a%402l/test/analyse-phys/main.py
% timing: https://gitlab.com/jetdynamics/njet-tools/-/blob/master/3g2a%402l/test/analyse-phys/timing.py

\begin{figure}
    \begin{center}
        \includegraphics[width=0.7\textwidth]{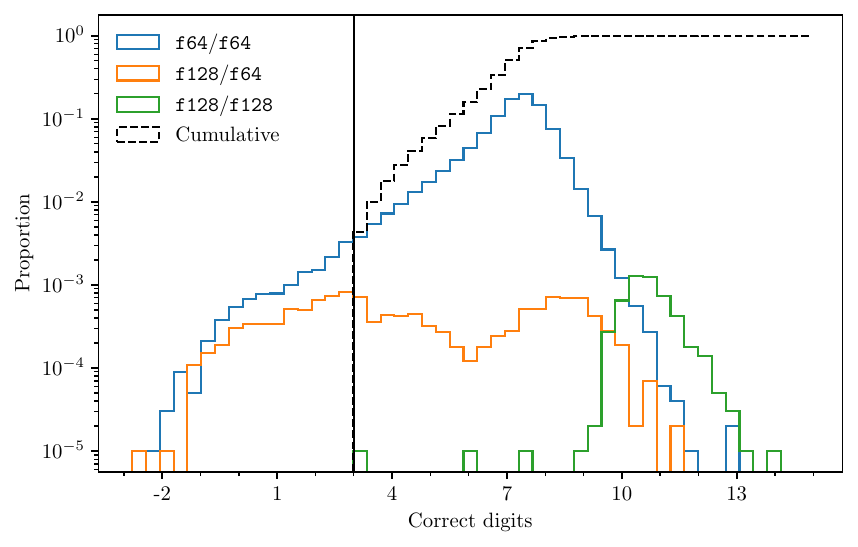}
        \caption{
            \label{fig:stability}
            Histogram of the error estimate of the $\mathcal{H}^{(2)}$ evaluations as given by the dimension scaling test.
            We use the evaluation strategy with a target accuracy of three digits, denoted by the vertical solid black line, and show errors for all precision levels as well as the cumulative error on all passing points.
            A cumulative bin of height $h$ at $d$ digits indicates $h$ proportion of points have an accuracy of at best $d$ digits.
        }
    \end{center}
\end{figure}

To assess the stability and speed of our implementation, we evaluate the hard function $\mathcal{H}^{(2)}$ over \SI{100}{\kilo\relax} points with a minimum target accuracy of three digits (corresponding to a maximum relative error of $10^{-3}$).
The phase-space sampling density is determined by the \acl{LO} process, obtained from \nnlojet~\cite{Gauld:2019ntd}, to ensure a realistic setup regarding Monte Carlo cross-section calculations.
We use \njet v3.1.1 with dependencies \eigen v3.4.0~\cite{!eigenweb}, \qd v2.3.23, and \pentagonfunctions v2.0.1.
We compile with \gcc v12.1.1 on Rocky Linux 8.7.
We run the test on a machine with dual Intel Xeon Gold 5218 \acsp{CPU} running at \SI{2.3}{\GHz} under full compute load over 64 Hyper-Thread cores.

To demonstrate the stability, we histogram the $\mathcal{H}^{(2)}$ errors in \cref{fig:stability}.
We see \SI{1.3}{\percent} of points failing \ac{f64}/\ac{f64} evaluation, with \SI{0.8}{\percent} passing at \ac{f128}/\ac{f64} and \SI{0.5}{\percent} passing at \ac{f128}/\ac{f128}.
There is a double-humped shape to the \ac{f128}/\ac{f64} histogram; since \ac{f128}/\ac{f128} reevaluations vastly improve the accuracy, the left hump appears to correlate with points that are limited in accuracy by the pentagon functions, showing that the stability bottleneck lies in the numerical evaluation of the special functions.
The evaluation strategy achieves the target accuracy for all points.

We find a single \ac{f64}/\ac{f64} call has a mean time of \SI{2.6}{\second} with \SI{96}{\percent} in evaluation of pentagon functions.
Using the evaluation strategy with three digit target accuracy, we obtain a mean timing per phase-space point of \SI{7.4}{\second} with \SI{99}{\percent} in evaluation of pentagon functions.
The bottleneck for evaluation speed is clearly also in the special functions.

These timing results are improved from the original tests reported in \incite{Badger:2021imn}.
Since then, we have made some improvements to \njet while implementing the five-parton scattering channels released in v3.1.0~\cite{Moodie:2022dxs}, are using newer development servers, and most significantly are using the updated version of \pentagonfunctions~\cite{Chicherin:2021dyp}.

\section{Infrared stability}
% https://gitlab.com/jetdynamics/njet-tools/-/tree/master/3g2a-2l-ir-perf

\begin{figure}
    \begin{center}
        \includegraphics[width=0.7\textwidth]{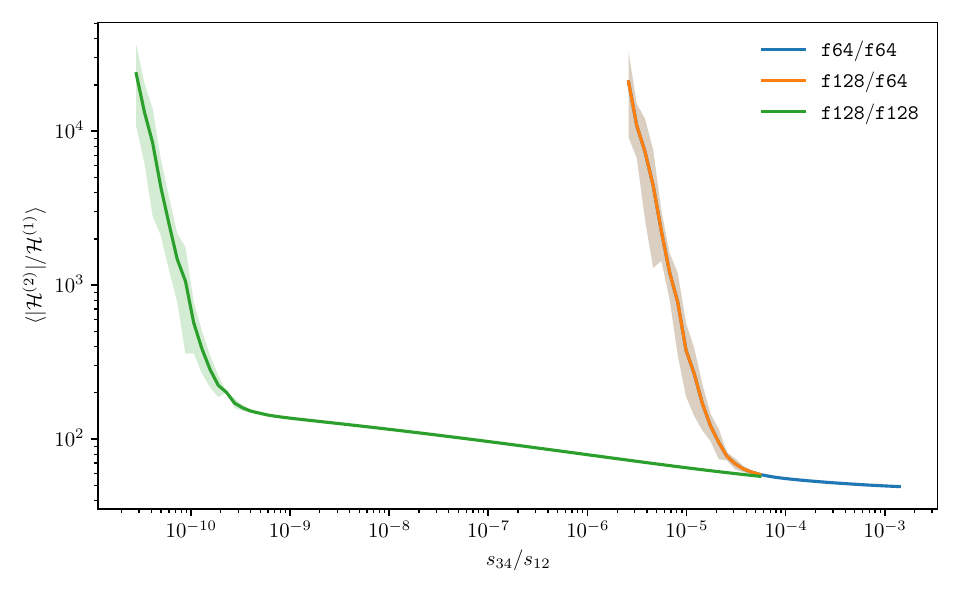}
        \caption{
            \label{fig:collinear}
            Plot of the value of $\mathcal{H}^{(2)}$ normalised by $\mathcal{H}^{(1)}$ and averaged over ten slices of phase space which drive towards a collinear limit in the first two outgoing legs.
            The error band is given by the dimension scaling test with $\mathcal{H}^{(1)}$ and $\mathcal{H}^{(2)}$ errors added in quadrature.
            Evaluations are made using the evaluation strategy with a target of three digits.
            Where it is not visible, the \ac{f64}/\ac{f64} (blue) line coincides with the \ac{f128}/\ac{f64} (orange) line.
        }
    \end{center}
\end{figure}

We prepare another phase space using the prescription of \incite{Badger:2015cxa}, generating a \num{100}-point slice that approaches an infrared limit.
The method generates a five-particle phase space in a collinear parametrisation from a four-particle phase space, described in \crefext{section}{2.4.1} of \incite{Moodie:2022dxs}.
We perform this for ten different four-particle seeds and plot the mean, to avoid any irregularities that may arise when approaching the limit in an exceptional direction, in \cref{fig:collinear}.
The lower precision evaluations diverge at around $s_{34}/s_{12}=10^{-4}$.
The origin of this numerical divergence lies in the evaluation of the pentagon functions as it is not fixed by the \ac{f128}/\ac{f64} evaluation, which agrees with the conclusion on stability bottlenecks in \cref{sec:performance}.
Evaluation in \ac{f128}/\ac{f128} remains unproblematic deep into the limit; it does not begin to diverge until below $s_{34}/s_{12}=10^{-9}$.

The same infrared stability test was performed on the leading-colour double-virtual amplitudes for \ac{NNLO} trijet production at hadron colliders, $pp\to jjj$~\cite{Moodie:2022dxs}.
Comparatively, the limit of stability is reached much sooner in the full-colour $gg\to g\gamma\gamma$ calculation, as is expected due to the higher complexity of the additional non-planar topologies.
However, if evaluations were required deeper into the infrared, the evaluation strategy could simply be extended to include \ac{f256} evaluations via \qd, although this would incur a large runtime penalty.

This demonstrates that the amplitudes are suitable not only for integrating over the two-to-three virtual phase space at \ac{NLO}, but also the more difficult two-to-two real-virtual phase space at \ac{NNLO}.
In fact, with libraries such as \njet and \openloops~\cite{Buccioni:2019sur,Badger:2021ohm} and the recent calculation of the full-colour three-loop $gg\to\gamma\gamma$ amplitudes~\cite{Bargiela:2021wuy}, all amplitude ingredients are now available for such a calculation.

\section{Conclusion}

We discussed the current state of full-colour two-to-three two-loop amplitudes in massless \acs{QCD}, focussing on the example of $gg\to g\gamma\gamma$.
We reviewed some optimisation strategies for the reconstruction of these amplitudes from numerical evaluations over finite fields.
Then we tested the performance of our \cpp library implementation, showing it to be fast and stable, even when pushed towards infrared limits.
The results demonstrate the readiness of this class of amplitudes to be used in precision studies in the percent-level era of phenomenology.

\ack{
    RM thanks Simon Badger and Simone Zoia for useful discussion and comments.
    This project received funding from the EU Horizon 2020 research and innovation programme \textit{High precision multi-jet dynamics at the LHC} (grant agreement No 772099).
}

\bibliography{proc,custom}

\end{document}